\documentclass[aps,prl,twocolumn,showpacs,floatfix]{revtex4}

\usepackage{amssymb}
\usepackage{amsmath}
\usepackage{graphicx}
\usepackage{amssymb}
\usepackage{amsfonts}

\usepackage{color}

\RequirePackage[hyperindex,colorlinks,bookmarksnumbered,plainpages=true]{hyperref}
\usepackage{hyperref}

\definecolor{green2}{rgb}{0,0.8,0}
\hypersetup{linkcolor=blue,urlcolor=blue,citecolor=green2}

\newcommand{\Eq}[1]{Eq.~(\ref{#1})}

\newcommand{\Fig}[1]{Fig.~\ref{#1}}
\newcommand{\Figs}[1]{Figs.~\ref{#1}}

\def\X{{\scriptstyle{\rm X}}}
\def\Xp{{\scriptstyle{\rm X}^\prime}}
\def\x{{\scriptscriptstyle{\rm X}}}
\def\xp{{\scriptscriptstyle {\rm X}^\prime}}

\def\t{{\scriptscriptstyle {\rm D}}}
\def\k{{\scriptscriptstyle {\rm K}}}
\def\T{{\scriptstyle {\rm D}}}
\def\K{{\scriptstyle {\rm K}}}

\def\sp{{s^\prime}}
\def\n0{{n_0}}
\def\om0{{\omega_0}}
\def\B{{{\cal B}}}
\def\C{{\cal C}}

\def\Mk{{M_{\rm K}}}
\def\Tk{{T_{\rm K}}}

\def\red{{\rm red}}
\def\raw{{\rm raw}}

\def\rhs{{RHS}}

\def\eg{\emph{e.g. }}
\def\ie{\emph{i.e. }}

\def\qph{\quad \phantom{.}}

\def\improved{{\rm im }}
\def\discarded{{discarded}}

\def\FL{{Fermi--liquid}}

\begin{document}

\preprint{ }

\title[ShortTitle]{ Sum-rule Conserving Spectral Functions from the
  Numerical Renormalization Group }
\author{Andreas Weichselbaum and Jan von Delft}
\affiliation{Physics Department, Arnold Sommerfeld Center for Theoretical
  Physics, and Center for NanoScience,
  Ludwig-Maximilians-Universit\"at M\"unchen, D-80333 M\"unchen,
  Germany}

\begin{abstract}

We show how spectral functions for quantum impurity models can be
calculated very accurately using a complete set of ``\discarded''
numerical renormalization group eigenstates, recently introduced by
Anders and Schiller.  The only approximation is to judiciously exploit
energy scale separation.  Our derivation avoids both the overcounting
ambiguities and the single-shell approximation for the equilibrium
density matrix prevalent in current methods, ensuring that relevant
sum rules hold rigorously and spectral features at energies below the
temperature can be described accurately.

\end{abstract}
\date[Date: ]{Jul 30, 2007}



\maketitle

Quantum impurity models describe a quantum system with a small
number of discrete states, the "impurity", coupled to a continuous
bath of fermionic or bosonic excitations. Such models are relevant
for describing transport through quantum dots, for the treatment of
correlated lattice models using dynamical mean field theory, or for
the modelling of the decoherence of qubits.

The impurity's dynamics in thermal equilibrium can be characterized by
spectral functions of the type
$ {\cal A}^{{\cal B}{\cal C}} (\omega) = \int \frac{dt}{2\pi} e^{i
  \omega t} \langle \hat \B (t) \hat \C \rangle_T$. Their Lehmann
representation reads
\begin{eqnarray}
  \label{eq:Lehmann}
  {\cal A}^{{\cal B}{\cal C}}
 (\omega) = \sum_{a,b}
  \langle b | \hat \C | a \rangle
  \frac{e^{- \beta E_a}}{Z}
  \langle a | \hat \B | b \rangle
\, \delta ( \omega - E_{ba}) \; ,
\end{eqnarray}
with $Z= \sum_a e^{- \beta E_a}$ and $E_{ba} = E_b \!-\!  E_a$, which
implies the sum rule $ \int d\omega {\cal A}^{{\cal B} {\cal C}}
(\omega) = \langle \hat \B \hat \C \rangle_T$.  In this Letter, we
describe a strategy for numerically calculating ${\cal A}^{{\cal B}
{\cal C}} (\omega)$ that, in contrast to previous methods,
rigorously satisfies this sum rule and accurately describes both high
\emph{and} low frequencies, including $\omega \lesssim T$, which we
test by checking our results against exact Fermi liquid relations.

Our work builds on  Wilson's numerical renormalization group (NRG)
method \cite{Wilson75}. Wilson
discretized the environmental spectrum on a logarithmic grid of
energies $\Lambda^{-n}$, (with $\Lambda> 1$, $1 \le n \le N \to
\infty$), with exponentially high resolution of low-energy
excitations, and mapped the impurity model onto
a ``Wilson tight-binding chain'', with hopping matrix elements that
decrease exponentially as $\Lambda^{-n/2}$ with site index $n$.
Because of this separation of energy scales, the Hamiltonian
can be diagonalized iteratively: adding one site at a
time, a new ``shell'' of eigenstates is constructed from the new
site's states and the $\Mk$ lowest-lying eigenstates of the
previous shell (the so-called ``kept'' states), while
``discarding'' the rest.

Subsequent authors
\cite{SakaiShimizuKasuya89,YoshidaWhitakerOiveira90,Costi94,Costi97,
BullaHewsonPruschke98,Bulla07,Hofstetter00,BullaCostiVollhardt01,
BullaTongVojta03}
have shown that spectral functions such as ${\cal A}^{{\cal B} {\cal
C}} (\omega)$ can be calculated via the Lehmann sum, using NRG states
(kept and \discarded) of those shells $n$ for which $\omega \sim
\Lambda^{-n/2}$.  Though plausible on heuristic grounds, this
strategy entails double-counting ambiguities \cite{Costi97} about how
to combine data from successive shells. Patching schemes
\cite{BullaCostiVollhardt01} for addressing such ambiguities involve
arbitrariness.  As a result, the relevant sum rule is not satisfied
rigorously, with typical errors of a few percent.  Also, the
density matrix (DM) $\hat \rho = e^{- \beta \hat H}/Z$ has hitherto
been represented rather crudely using only the single $N_T$-th shell
for which $T \simeq \Lambda^{-\frac{1}{2}(N_T-1)}$
\cite{Hofstetter00}, with a chain of length $N=N_T$, resulting in
inaccurate spectral information for $\omega \lesssim T$.  In this
Letter we avoid these problems by using in the Lehmann sum an
approximate but \emph{complete} set of eigenstates of the
Hamiltonian, introduced recently by Anders and Schiller (AS)
\cite{AndersSchiller05}.

\begin{figure}[tb]
\includegraphics[width=1\linewidth]{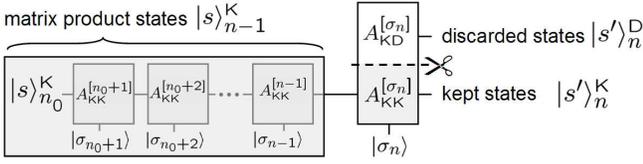}
\caption{
   Diagram for the kept (or \discarded) matrix product state
   $|s^\prime \rangle^\k_n$ (or $|s^\prime \rangle^\t_n$): the
   $n$th box represents the matrix block $A^{[\sigma_n]}_{\k \x}$,
   its left, bottom and right legs carry the labels of the states $|
   s\rangle_{n-1}^\k$, $|\sigma_{n}\rangle $ and
   $|s'\rangle_n^{\k} $ (or $|s'\rangle_n^{\t} $), respectively.
}
\vspace{-0.1in} \label{fig:mps}
\end{figure}

\emph{Wilson's truncation scheme.---}
 The Wilson chain's zeroth site represents the bare impurity
Hamiltonian $\hat h_0$ with a set of $d_0$ impurity
states $|\sigma_0\rangle$. It is
coupled to a fermionic chain, whose $n$th site ($1 \le
n \le N$) represents a set of $d$ states $|\sigma_n\rangle$,
responsible for providing energy resolution to the spectrum at the
scale $\Lambda^{-n/2}$.  For a spinful fermionic band, for example,
$\sigma_n \in \{ 0, \uparrow, \downarrow, \uparrow \downarrow \}$,
hence $d=4$. (Bosonic chains can be treated similarly
\cite{BullaTongVojta03}.)  The Hamiltonian $\hat H = \hat H_N$ for the
full chain is constructed iteratively by adding one site at a time,
using $\hat H_n = \hat H_{n-1} + \hat h_n$ (acting in a $d^n
d_0$-dimensional Fock space ${\cal F}_n$ spanned by the basis states
$\{ | \sigma_n\rangle \otimes \dots \otimes |\sigma_0\rangle \}$),
where $\hat h_n$ links sites $n$ and $n-1$ with hopping strength
$\sim\!\!\Lambda^{-n/2}$.
Since the number of eigenstates of $\hat H_n$ grows exponentially
with $n$, Wilson proposed the following iterative truncation
scheme to numerically diagonalize the Hamiltonian:
Let $\n0$
be the last iteration for which a complete set
$\{ |s \rangle_\n0^\k \}$ of ``kept'' eigenstates of $\hat H_{\n0}$
can be calculated without trunction.
For $n > \n0$, construct the orthonormal
eigenstates $\{ |s \rangle_n^\x \}$ of $\hat H_n$ (the $n$th
``shell''), with eigenvalues $E^n_s$, as linear combinations of the
kept eigenstates $|s\rangle_{n-1}^\k $ of $\hat H_{n-1}$ and the
states $|\sigma_n\rangle$ of site $n$,
\begin{equation}
  |s^\prime \rangle_n^{\x} = \sum_{\sigma_n s}^\k
  |\sigma_n \rangle \otimes |s \rangle_{n-1}^\k \;
  \bigl[A_{\k\x}^{[\sigma_n]}\bigr]_{s s^\prime} \; ,
\label{eq:newsAolds}
\end{equation}
with coefficients arranged into a matrix $A^{[\sigma_n]}_{\k\x}$ whose
elements are labelled by $ss'$.  The superscript $\X\!=\!\K$ or $\T$
indicates that the new shell has been partitioned into ``kept'' states
(say the $\Mk$ lowest-lying eigenstates of $\hat H_n$) to be retained
for the next iteration, and ``\discarded'' states (the remaining
ones).  Since $\hat h_n$ acts as a weak perturbation (of relative size
$\Lambda^{-1/2}$) on $\hat H_{n-1}$, the $d$-fold degeneracy of the
states $|\sigma_n \rangle \otimes |s \rangle_{n-1}^\x$ is lifted,
resulting in a characteristic energy spacing $\Lambda^{-n/2}$ for
shell $n$.  Iterating until the spectrum of low-lying eigenvalues has
reached a fixed point (for $n=N$, say), one generates a set of
eigenstates $\{ |s\rangle_n^\x \}$ with the structure of matrix
product states \cite{VWSCvD05} (\Fig{fig:mps}).  The states
generated for the last, $N$th shell will all be regarded as
``\discarded'' \cite{AndersSchiller05}.

\emph{Anders-Schiller basis.---}
Recently, AS have shown \cite{AndersSchiller05} that the \discarded\
states can be used to build a complete basis for the whole Wilson
chain: the states $\{| s \rangle_n^\x \}$ describing the $n$th shell
are supplemented by a set of $d^{N-n}$ \emph{degenerate}
``environmental'' states $\{ |e_n\rangle = | \sigma_N \rangle \otimes
\dots \otimes | \sigma_{n+1} \rangle \}$ spanning the rest of the
chain ($|e_N\rangle$ denotes an ``empty'' state),
to construct the set of states $\{|se \rangle^\x_n \equiv |e_n
\rangle \otimes |s \rangle_n^\x \}$. These reside in the complete Fock
space ${\cal F}_N$ of the full chain, spanning ${\cal F}_N$ if $n
\leq \n0$.
Ignoring the degeneracy-lifting effect of the rest of the
chain, these states become approximate eigenstates of the
Hamiltonian $\hat H_N$ of the \emph{full} chain
(``NRG-approximation''),
\begin{equation}
  \hat H_N |s e\rangle^\x_n  \simeq E_s^{n}
  |s e\rangle_n^\x  \; ,
  \label{eq:NRGapprox}
\end{equation}
with eigenenergies \emph{independent} of the $(d^{N-n})$-fold
degenerate environmental index $e_n$. (This  will
facilitate tracing out the environment below).  By construction, we
have ${}_m^{\hspace{0.3mm} \t} \hspace{-0.5mm} \langle se| \sp
e^\prime \rangle_{n}^\t =\delta_{mn}\delta_{e^{\vphantom{\prime}}_{n}
  e^\prime_{n}} \delta_{s\sp}$ and
\begin{equation}
{}_m^\k \hspace{-0.5mm}
\langle  s e | \sp e^\prime \rangle_{n}^\t = \left\{ \!
\begin{array}{ll}
    0,  &  m\ge n \\
    \delta_{e^{\vphantom{\prime}}_{n} e^\prime_{n}}
    \left[A^{\sigma_{m+1}}_{\K\K} \dots
          A^{\sigma_{n \vphantom{+1}}}_{\K\T} \right]_{s\sp},
    & m<n.
\end{array}
\right. \label{subeq:tt-tk-orthonormality}
\end{equation}
The \discarded\ states of shell $n$ are orthogonal to the \discarded\
states of any other shell, and to the kept states of that or any later
shell. Combining the \discarded\ states from all shells thus yields a
complete set of NRG eigenstates of $\hat H_N$,
the ``Anders-Schiller basis'', that span the full Fock
space ${\cal F}_N$ ($\sum_n$ henceforth stands
for $\sum_{n > \n0}^N$):
\begin{eqnarray}
  \label{eq:identity}
  \boldsymbol{1}^{(d_0 d^{N})} =
  \sum_{s e}
  |s e\rangle_{\n0}^\k \;
  {}_{\n0}^{\, \hspace{0.61mm} \k} \hspace{-0.3mm} \langle se |
=
  \sum_n
  \sum_{se}
  |se \rangle_n^\t \;
  {}_n^{\hspace{0.0mm} \t} 
\langle s e | \,  .
\end{eqnarray}

\emph{Local Operators.---} Let us now consider a ``local'' operator $\hat \B$
acting nontrivially only on sites up to $\n0$. Two particularly useful
representations are
\begin{eqnarray}
  \label{eq:OpRep}
\hat \B =
\sum_{s \sp e}
|se\rangle_\n0^\k
\; [\B_{\k \k}^{[\n0]}]_{s \sp}^{\,} \;
  {}_{\n0}^{\, \hspace{0.61mm} \k} \hspace{-0.3mm}
\langle \sp e |
 = \sum_n
\sum_{\x \xp}^{\neq \k \k}   \hat \B^{[n]}_{\x \xp} \, .
\end{eqnarray}
The left equality, written $\hat {\cal B} = \hat \B^{[\n0]}_{\k \k}$
in brief, represents the operator in the complete basis set $\{ | s
e\rangle_{\n0}^\k \}$, with matrix elements known exactly numerically
(possibly up to fermionic minus signs depending on the environmental
states; but these enter quadratically in correlation functions and
hence cancel).  The right-hand side (\rhs) of \Eq{eq:OpRep}
expresses $\hat {\cal B}$ in the AS-basis
and is obtained as follows: starting from $\hat
\B^{[\n0]}_{\k \k}$, one iteratively refines the ``kept-kept'' part of
$\hat {\cal B}$ from, say, the ($n-1$)st iteration in terms of the NRG
eigenstates $\{ |se\rangle^\x_n \}$ of the next shell, including both
kept and \discarded\ states ($\X=\K,\T$),
\begin{eqnarray}
  \label{eq:Oniterate}
  \hat \B_{\k \k}^{[n-1]} =
   \sum_{\x \xp} \sum_{ s \sp e }
  |s e \rangle_n^\x  \; \bigl[\B^{[n]}_{\x \xp}\bigr]_{s \sp}^{\,} \;
  {}_n^{\xp} \hspace{-0.7mm}
  \langle \sp e | =
  \sum_{\x \xp} \hat \B^{[n]}_{\x \xp} \, , \qph
\end{eqnarray}
thereby defining the operators $\hat \B^{[n]}_{\x \xp}$,
with  matrix elements
$\bigl[\B^{[n]}_{\x \xp}\bigr]_{s \sp}^{\,} = \bigl[A^{[\sigma_n
]\dagger}_{\x \k}\B_{\k \k}^{[n-1]} A^{[\sigma_n ]}_{\k \xp} \bigr]_{s \sp}$.
Splitting off all $\X \Xp \neq \K \K$ terms $(\T \T, \K \T,\T \K)$
and iteratively refining each
$\K \K$ term until $n=N$, we obtain the \rhs\ of \Eq{eq:OpRep}.
It has two important features: First, the matrix elements of the
time-dependent operator $\hat \B (t) = e^{i \hat H t} \hat \B e^{-i
\hat H t}$, evaluated within the NRG approximation, $[\B_n^{\x
\xp}(t)]_{s \sp}^{\,} \simeq [\B_n^{\x \xp}]_{s \sp}^{\,} \, e^{i t (
E_s^n - E_{\sp}^n)}$, contain differences of eigenenergies from the
\emph{same} shell only, \ie calculated with the same level of
accuracy.  Second, by \emph{excluding} $\K\K$ terms it rigorously
avoids the double-counting ambiguities 
and heuristic patching rules plaguing previous approaches
\cite{YoshidaWhitakerOiveira90,SakaiShimizuKasuya89,Costi94,%
Costi97,BullaHewsonPruschke98,Bulla07,BullaCostiVollhardt01,BullaTongVojta03,%
Hofstetter00}.

\emph{Thermal averages.---} To calculate thermal averages $\langle
\dots \rangle_T = {\rm Tr}[\hat \rho \dots ]$, we write the
full density matrix $\hat \rho = e^{-\beta \hat H}/Z $ using the
NRG approximation as in \Eq{eq:NRGapprox}
\begin{equation}
  \label{eq:rhont}
\hat{\rho} \simeq \sum_{n}
\sum_{se}
|se \rangle_n^\t \:
\frac{e^{- \beta E_{s}^n}}{Z} \:
    {}_n^{\t} \hspace{-0.2mm}
\langle se | \; = \sum_{n}
w_n \, \hat \rho^{[n]}_{\t \t}
\; ,
\end{equation}
where $w_n\!\equiv\!d^{N-n} Z^\t_n/Z$ and $Z_n^\t \!\equiv\! \sum_s^\t
e^{- \beta E_s^n}$. The
\rhs\ of \Eq{eq:rhont} expresses $\hat \rho$ as sum over
$\hat \rho^{[n]}_{\t \t}$, the density matrix for the \emph{\discarded}
states of shell $n$, properly normalized as $\mathrm{Tr}\,[\hat
\rho^{[n]}_{\t \t}]=1$, and
entering with relative weight $w_n$, with $\sum_n w_n = 1$.
Similarly, for spectral functions we have
\begin{eqnarray}
  \label{eq:GAassumsonm}
  \langle \dots \rangle_T  =
  \sum_n
  w_n \, \langle \dots  \rangle_n \; , \quad
  {\cal A} (\omega) = \sum_n
  w_n \, {\cal A}_n (\omega) \; , \qph
\end{eqnarray}
where the averages $\langle \dots \rangle_n$ and spectral functions
${\cal A}_n (\omega)$ are calculated with respect to $\hat
\rho^{[n]}_{\t \t}$ of shell $n$ only.

\begin{figure}[t]
   \includegraphics[width=1\linewidth]{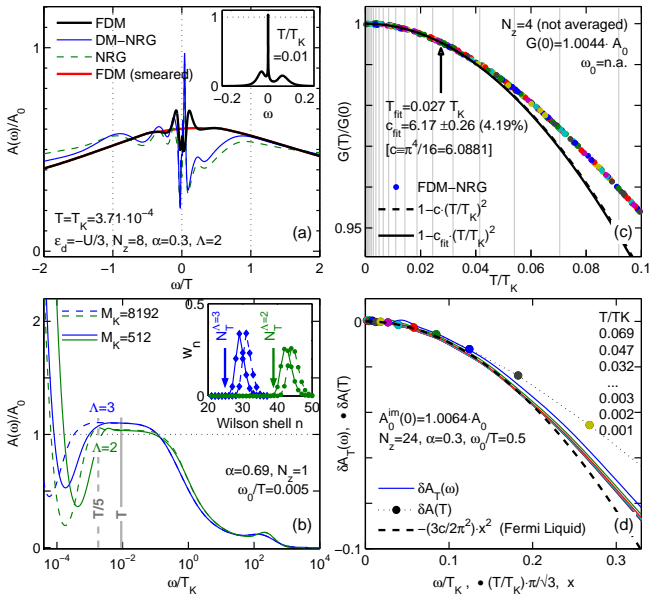}
\vspace{-0.3in}
\caption{ (color online). FDM-NRG results for the spectral function
  ${\cal A}_T (\omega)$ of the SIAM, with $U=0.12$, $\Gamma=0.01$,
  $\epsilon_d=-U/2$ ($\Tk=2.185\,10^{-4}$), $\Lambda=1.7$ and
  $\Mk=1024$, unless indicated otherwise.  Inset of (a): FDM-NRG
  result for $A_T (\omega)$ with $\omega$ in units of bandwith.  For
  (a),(b) an unconventionally small smearing parameter was used, $\om0
  =0.005\,T$ (except for thick gray (red) curve in (a), with $\om0 = 0.5\,T$),
  leading to spurious low-frequency oscillations.  These illustrate
  the differences (a) between NRG (dashed green curve),
  DM-NRG (solid thin blue curve) and FDM-NRG (black curve) results
  for the regime $\omega \lesssim T $; and (b) between
  different choices of $\Mk$ and $\Lambda$ for FDM-NRG, which yield
  different shapes for the weigths $w_n$ [shown in inset of (b)]:
  larger $\Lambda$ reduces the scale $\delta_T$ at which
  oscillations set in, but yields less accurate values for Kondo
  peak height in the regime $\delta_T \lesssim \omega \lesssim \Tk$.
  (c),(d) Comparison of high quality FDM-NRG data (dots, solid curves)
  with exact \FL\ results (black dashed lines) for (c) the conductance
  $G(T)$ for $T \ll \Tk$; and (d) for ${\cal A}^\improved_T (\omega)$
  for $T, \omega \ll \Tk$.
  In (c), $c_{\rm fit}$ was found from a data fit to $c_{\rm fit}
  (T/\Tk)^2$ for $T< T_{\rm fit}$ (arrow).  In (d) we plot
  $\delta {\cal A}_T (\omega) =
  [A^\improved_T(\omega) - A^\improved_T(0)]/A^\improved_0(0)$
  vs. $\omega/\Tk$ (curves) and
  $\delta {\cal A}(T) = [A^\improved_T(0)/ A^\improved_0 (0) - 1]$
  vs. $(T/\Tk) \pi /\sqrt 3$
  (dots), for a set of 12 temperatures between 0.001 and 0.069 $\Tk$
  (with curves and dots having same $T$ in the same color), to
  illustrate the leading $\omega$- and $T$-behavior of ${\cal
  A}_T^\improved (\omega)$; the dashed black line represents the
  expected Fermi liquid behavior in both cases,
  $-\frac{3c}{2\pi^2} x^2$ vs. $x$.
}
\vspace{-0.15in}\label{fig:data}
\end{figure}

Previous strategies
\cite{Costi94,Costi97,BullaHewsonPruschke98,Bulla07,
  Hofstetter00,BullaCostiVollhardt01,
  BullaTongVojta03,AndersSchiller05} for thermal averaging amount to
using a ``single-shell approximation'' $w_n = \delta_{n N_T}$ for the
density matrix and terminating the chain at a length $N=N_T$ set by $T
\simeq \Lambda^{-\frac{1}{2}(N_T-1)}$.  As a result, spectral features
on scales $\omega\le T$, which would require a longer chain, are
described less accurately [see \Figs{fig:data}(a) and \ref{fig:data}(b)].
Our novel approach avoids
these problems by using the \emph{full} density matrix (FDM), summed
over \emph{all} shells, letting the weighting function $w_n$ select
the shells relevant for a given temperature yielding a
smooth $T$ dependence [see \Fig{fig:data}(c)].  Since $w_n$ has a peak
width of five to ten shells depending on $\Lambda$, $d$ and $\Mk$
and peaks at $n$-values somewhat above $N_T$ [arrow
\Fig{fig:data}(b)], spectral information from energies well below $T$
is retained.

Let us now consider the spectral function $ \cal{A}^{\B\C}
(\omega)$, for local operators $\hat\B$ and $\hat\C$.
Equations (\ref{subeq:tt-tk-orthonormality}),
(\ref{eq:OpRep}), (\ref{eq:rhont}), and (\ref{eq:GAassumsonm})
can be used to evaluate $
\langle \hat {\cal B} (t) \hat {\cal C} \rangle_n$. Fourier
transforming the result we find (sums over $s \sp$ and $\sigma_n$
implied):
\begin{eqnarray}
  & & \phantom{.} \hspace{-5mm}
  {\cal A}^{{\cal B} {\cal C}}_n (\omega)
  =  \!\! \sum_{m > \n0 }^n
  \sum_{\x \xp }^{\neq \k \k}
  \left[{\cal C}^{[m]}_{\xp \x} \,
    \rho^{[mn]}_{\x^{\vphantom{\prime}}
      \x}\right]_{\sp s  }^{\,}  \!\!
  \left[{\cal B}_{\x \xp}^{[m]}\right]_{s \sp}^{\,} \!\!
  \delta( \omega - E_{\sp s}^m ) \; , \nonumber
\\ \label{eq:correlation}
\label{eq:rhored}
&&  \phantom{.} \hspace{-6mm}
\left[\rho^{[m=n]}_{\t \t} \right]_{s \sp}  \!\! =
\delta_{s \sp} \, \frac{e^{- \beta E_s^n}}{Z_n} \; ,
\rule[-5mm]{0mm}{0mm}
\\  &&  \phantom{.} \hspace{-6mm}
\left[\rho^{[m<n]}_{\k \k} \right]_{s \sp} \!\! =
 \left[ A^{[\sigma_{m+1}]}_{\k \k}
  \dots A^{[\sigma_n]}_{\k \t}
\rho^{[nn]}_{\t \t}
A^{[\sigma_{n}] \dagger}_{\t \k}  \!
  \dots A^{[\sigma_{m+1}] \dagger}_{\k \k} \right]_{
  s \sp} \!\! .
\nonumber
\end{eqnarray}
Similarly, the static quantity $\langle \hat {\cal B} \hat {\cal C}
\rangle_n$ equals the first line's \rhs\ without the
$\delta$~function. The matrix elements $\bigl[\rho^{[mn]}_{\x \x}
\bigr]_{s \sp}^{\,} \equiv \sum_{e}\ {}_m^\x \langle s e | \hat
\rho^{[n]}_{\t \t} | \sp e \rangle_m^\x $ are given by the second and
third lines, together with $\rho^{[m=n]}_{\k \k} = \rho^{[m<n]}_{\t
  \t} = 0$.  After performing a ``forward run'' to generate all
relevant NRG eigenenergies and matrix elements, ${\cal A}^{{\cal B}
  {\cal C}} (\omega)$ can be calculated in a single ``backward run'',
performing a sum with the structure $\sum_{m>n_0}^N [{\cal C}
\rho^\red {\cal B} \cdot \delta()]^{[m]}$, starting from $m=N$. Here $
\rho_{\x\x}^{[m],\red} \equiv \sum_{n \ge m}^N w_n \rho^{[mn]}_{\x
  \x}$ (updated one site at a time during the backward run) is the
\emph{full} reduced density matrix for shell $m$, obtained iteratively
by tracing out all shells at smaller scales $\Lambda^{-n/2}$ ($n\ge
m$).

Equations (\ref{eq:rhont})--(\ref{eq:rhored}) are the main results of
our ``FDM-NRG'' approach.
They rigorously generalize
Hofstetter's DM-NRG \cite{Hofstetter00} (which leads
to similar expressions, but using $w_n = \delta_{n N_T}$ and without
excluding $\K \K$ matrix elements), and
provide a concise prescription, free from
double counting ambiguities, for how to combine NRG data from
different shells when calculating ${\cal A}^{{\cal B} {\cal
   C}}(\omega)$.
The relevant sum rule is satisfied
\emph{identically}, since by construction $\int d \omega {\cal
  A}^{{\cal B} {\cal C}}_n (\omega) = \langle \hat {\cal B} \hat {\cal
  C} \rangle_n$ holds for every $n$ and arbitrary temperature and
NRG-parameters $\Lambda$ and $\Mk$.

\emph{Smoothing discrete data.---} We obtain smooth curves for ${\cal
A}^{{\cal B} {\cal C}} (\omega)$ by broadening the discrete
$\delta$~functions in \Eq{eq:correlation} using a broadening kernel
that smoothly interpolates from a log-Gaussian form (of width
$\alpha$) \cite{SakaiShimizuKasuya89,Costi94} for $|\omega| \gtrsim
\om0$, to a regular Gaussian (of width $\om0$) for $|\omega| <\om0$,
where $\om0$ is a ``smearing parameter'' whose significance is
explained below.  To obtain high quality data, we combine small
choices of $\alpha$ with an average over $N_z$ slightly shifted
discretizations \cite{YoshidaWhitakerOiveira90} (see \cite{app} for
more details).

\emph{Application to Anderson model.---} We illustrate our method for
the standard single-impurity Anderson model (SIAM).  Its local
Hamiltonian $\hat h_0 \equiv \sum_\sigma \epsilon_0
c_{0\sigma}^{\dagger} c_{0\sigma}^{\phantom{\dagger}}+ U
c_{0\uparrow}^{\dagger} c_{0\uparrow}^{\phantom{\dagger}}
c_{0\downarrow}^{\dagger} c_{0\downarrow}^{\phantom{\dagger}} $
describes a localized state with energy $\epsilon_{0}$, with a Coulomb
penalty $U$ for double occupancy.  It is coupled to a Wilson chain
$\sum_{n \sigma} \lambda_n (c^\dagger_{n+1 \sigma}
c^{\vphantom{\dagger}}_{n\sigma} + {\rm h.c.})$, which generates a
local level width $\Gamma$.  We calculated ${\cal A}^< (\omega)\equiv
{\cal A}^{c^\dagger_{0\sigma} \! c^{\vphantom{\dagger}}_{0\sigma}}
(-\omega)$, ${\cal A}^> (\omega) \equiv {\cal
  A}^{c^{\vphantom{\dagger}}_{0\sigma} c^{{\dagger}}_{0\sigma}}
(\omega)$ and ${\cal A} \equiv {\cal A}^> + {\cal A}^<$.  An
``improved'' version ${\cal A}^\improved $ thereof can be obtained by
calculating the impurity self energy $\Sigma(\omega,T)$
\cite{BullaHewsonPruschke98,app} via FDM-NRG, which is less sensitive
to smoothening details and yields more accurate results for the Kondo
peak height ${\cal A}_{T\simeq 0} (0)$ at zero temperature.

\emph{Sum rules.---} As expected, we find FDM-NRG to be significantly
more accurate at lower computational cost than NRG or DM-NRG
\cite{Hofstetter00,PetersPruschkeAnders06}. The sum rules
\begin{eqnarray}
  \label{eq:A=1}
  \int d \omega
  {\cal A}^{c^\dagger_{0\sigma} \! c^{\vphantom{\dagger}}_{0\sigma}}
  (\omega)
  = \langle {c^\dagger_{0\sigma} \!
    c^{\vphantom{\dagger}}_{0\sigma}}\rangle_T
  , \; \;   \int  d \omega {\cal A} (\omega) = 1  \; ,
\end{eqnarray}
hold exactly to $10^{-15}$ for our discrete data, and to $10^{-4}$
after smoothing (due to numerical integration inaccuracies).
Moreover, even for $\Mk$ as small as 256, our results for ${\cal
  A}_{T\simeq 0} (0)$ and ${\cal A}^\improved_{T \simeq 0} (0)$
typically agree to within 2\% and 0.2\%, respectively, with the
Friedel sum rule, which requires $\pi \Gamma {\cal A}^{\rm
  exact}_{T=0} = \sin^2 \pi \langle {c^\dagger_{0\sigma} \!
  c^{\vphantom{\dagger}}_{0\sigma}}\rangle_0$.  The exact relation $
{\cal A}^{<} (\omega) = f(\omega) {\cal A} (\omega)$ ($f$ is the Fermi
function), which follows from detailed balance, is likewise satisfied
well (though not rigorously so): the left-hand side of \Eq{eq:A=1}
typically equals $\int d \omega f (\omega) {\cal A} (\omega) $ to
better than $10^{-4}$.

\emph{Low-frequency data.---} Due to the underlying logarithmic
discretization, all NRG-based schemes for calculating
finite-temperature spectral functions inevitably produce spurious
oscillations at very low frequencies $|\omega|\ll T$. The scale
$\delta_T$ at which these set in can be understood as follows: the
Lehmann sum in \Eq{eq:Lehmann} is dominated by contributions from
initial states $|a\rangle$ with energy $E_a \simeq T$, represented by
NRG shells with $n$ near $N_T$. The characteristic energy scale of
these states limits the accuracy obtainable for energy differences
$E_{ba}$ to accessible final states $|b\rangle$. Thus the scale
$\delta_T$ is set by those shells which contribute with largest weight
$w_n$ to the density matrix.

We analyze this in more detail in \Figs{fig:data}(a) and
\ref{fig:data}(b) by purposefully
choosing the smearing parameter to be unconventionally small, $\om0
\ll T$. The resulting spurious oscillations are usually smeared out
using $\om0 \gtrsim \delta_T$ [\Fig{fig:data}(a), thick gray (red)
curve], resulting in quantitatively accurate spectral functions only
for $|\omega| \gtrsim \om0\simeq\delta_T$. For conventional NRG
approaches, the ``single-shell'' approximation $w_n = \delta_{nN_T}$
typically leads to $\delta_T \simeq T$, as can be seen in
\Fig{fig:data}(a) [dashed (green) and thin solid (blue) line].
In contrast, FDM-NRG yields
a significantly reduced value of $\delta_T \simeq T/5$
[\Fig{fig:data}(a), black line, and \Fig{fig:data}(b)], since the
weighting functions $w_n$ [inset of \Fig{fig:data}(b)] retain weight
over several shells below $N_T$, so that lower-frequency information
is included.

\emph{\FL\ relations.---} To illustrate the accuracy of our
low-frequency results, we calculated ${\cal A}^\improved_T(\omega)$
for $\omega, T \ll \Tk$ for the symmetric SIAM, and made quantitative
comparisons to the exact Fermi-liquid relations \cite{Hewson93}
\begin{eqnarray}
  &&\hspace{-0.4in} 
  {
    A_T ( \omega ) \simeq
    A_0 \left [ 1- \frac{c}{2}\bigl(\frac{T}{\Tk}\bigr)^{2} -
      \frac{3c }{2\pi^2}\bigl(\frac{\omega}{\Tk}\bigr)^{2} \right]
  } , \label{eq:FermiLiquid}
  \\ &&\hspace{-0.4in} 
  {
    G\left( T\right) \equiv \int\limits_{-\infty }^{\infty } \!\!
  d \omega \, A\left( \omega
      ,T\right) \bigl( -\frac{\partial f}{\partial \omega }\bigr)
    \simeq A_0\Bigl[ 1 - c\bigl(\frac{T}{\Tk}\bigr)^2\Bigr]
  } . \label{eq:conductance}
\end{eqnarray}
Here $A_0\!\equiv\!1/\pi\Gamma$, $c\!\equiv\!\pi^4/16$,
and the Kondo temperature $\Tk$ is defined via the
static magnetic susceptibility \cite{Costi94} $\chi_0 \bigl.\bigr|_{T=0} \equiv
1/4T_{K}$. Figures \ref{fig:data}(c) and \ref{fig:data}(d) show the FDM-NRG data
[gray (colored) dots and lines] to be in remarkable quantitative
agreement with these relations [black dashed curves]. The results for
the ``conductance'' $G(T)$, being a frequency integrated quantity
obtained by summing over discrete data directly without the need for
broadening, are more accurate than for ${\cal A}_T^\improved
(\omega)$, and reproduce the prefactor $c$ with an error consistently
below 5\% (hitherto, errors of order 10-30\% had been customary).
The smoothness of the data in \Fig{fig:data}(c), obtained using
temperatures not confined to the logarithmic set $\Lambda^{-n/2}$
[gray vertical lines in \Fig{fig:data}(c)], together with
the remarkable stability with respect
to different $z$-shifts illustrate the accuracy of our approach.

\emph{Conclusions.---} Our FDM-NRG method offers a transparent
framework for the calculation of spectral functions of quantum
impurity models, with much improved accuracy at reduced
computational cost. Its results satisfy frequency sum rules
rigorously and give excellent agreement with other consistency
checks such as the Friedel sum rule, detailed balance, or
Fermi-liquid relations, including the regime $\omega \lesssim T$.

We thank F. Anders, R. Bulla, T. Costi, T. Hecht, W. Hofstetter A.
Rosch and G. Z\'arand for discussions, and the KITP in Santa Barbara
for its hospitality.  The work was supported by DFG (SFB 631,
De-730/3-1,3-2), and in part by the NSF (PHY99-07949).

\emph{Note added.---} Just before
completion of this work we learned that Peters, Pruschke and Anders
had followed up on the same idea \cite{PetersPruschkeAnders06}.



\section{Appendix }

\emph{Smoothing Discrete Data:---}
The FDM-NRG yields the spectral function in the form of a Lehmann sum
over discrete $\delta$-functions, which have to be broadened to obtain
a smooth function.  Ideally, this should be done using a procedure for
which  the results are independent both of the parameters used for
broadening and for discretizing the conduction band.

We calculate the smoothened spectral function using ${\cal A}  \left(
  \omega\right) \equiv\int d\omega^{\prime} K \left( \omega
  ,\omega^{\prime}\right) {\cal A}_{\mathrm{raw}}\left(
  \omega^{\prime}\right) $.  Here ${\cal A}_\raw (\omega^\prime) = \sum_n
w_n {\cal A}_n (\omega^\prime)$ represents the discrete numerical data
obtained from \Eq{eq:rhored}, which in practice we collect in binned
form, typically using 250 bins per decade in $\omega^\prime$, so that $\int
d \omega^\prime$ becomes a sum over bins.  The raw data is folded with the
broadening kernel $K (\omega, \omega^\prime)$, which we choose to be of the
following form:
\begin{subequations}
\begin{eqnarray}
  \label{eq:gtot}
  K (\omega, \omega^\prime) & = &
  L (\omega, \omega^\prime) h ({\omega^\prime}) +
  G (\omega, \omega^\prime) [1 - h({\omega^\prime})]\hspace{0.2in}
\end{eqnarray}
where
\begin{eqnarray}
  \label{eq:gtot_sub}
  L (\omega,\omega^{\prime})   &  = &
  \textstyle{
     \frac{\theta ( \omega \omega^\prime)  }{\sqrt{\pi} \alpha |\omega|}
     e^{-\bigl( \frac{ \log | \omega /\omega^{\prime} | }
                     {\alpha}-\gamma\bigr)^{2}} \;
  }
\label{g(om,om')1}
\\
& = &
  \textstyle{
      \frac{\theta ( \omega \omega^{\prime})  }{\sqrt{\pi}
      \alpha |\omega^{\prime}| }
    e^{-\bigl( \frac{ \log | \omega^{\prime} /\omega |}
                    {\alpha} + \gamma - \frac{\alpha}{2} \bigr)^{2}}
    e^{-\alpha (\gamma-\frac{\alpha}{4})}\; ,
  } \nonumber \label{g(om,om')2}
\\
G (\omega,\omega^{\prime})   &  = &
  \textstyle{
     \frac{1 }{\sqrt{\pi} \omega_0} e^{
          -\bigl(\frac{\omega - \omega^\prime}{\omega_0}\bigr)^2
        } \; ,
  }
\\
h ({\omega^\prime})  & = &  \left\{ \begin{array}{ll}
    1 \; ,  & \quad |\omega^\prime| \ge \omega_0 \; , \\
    \textstyle{
       e^{-\bigl( \frac{ \log |\omega^\prime /\omega_0 | }{\alpha }
       \vphantom{+ \gamma/2} \bigr)^{2}}
    }  \; , &
       \quad |\omega^\prime| < \omega_0 \; .
 \end{array}
\right. \label{h(om)}
\end{eqnarray}
\end{subequations}
The chosen kernel $K$ constitutes a smooth interpolation, of somewhat
arbitrary shape $h({\omega^\prime})$, between a log-Gaussian
\cite{SakaiShimizuKasuya89,Costi94} broadening kernel $L$
on the one hand, used for all $\omega^\prime$-frequencies but the smallest
(with $ \omega$ and $\omega^\prime$ restricted to have the same sign); and a
Gaussian broadening kernel $G$ of width $\omega_0$ on the other, used
for $|\omega^\prime| < \omega_0$ to smoothly connect the regimes of positive
and negative frequencies.  We choose $\omega_0$ to be roughly a factor
of 2 smaller than the smallest energy scale in the problem, including
the Kondo temperature $T_K$ (note that by construction in \Eq{h(om)}
the transition to regular Gaussian sets in \emph{below} $\omega_0$).

The log-Gaussian kernel $L(\omega, \omega^\prime)$ was purposefully chosen
to have the following three desirable features: \\
(i) \emph{Frequency-dependent width:} being Gaussian on a logarithmic
scale, on a linear scale its width as function of $\omega$ is
proportional to $\omega^\prime$.  This is needed to deal with the fact that
spectral data generated using Wilson's logarithmic discretization grid
is more coarse-grained at large frequencies than at smaller ones. \\
(ii) \emph{Conservation of weight:} we have $\int d \omega L(\omega,
\omega^\prime) = 1$, ensuring that $\int d \omega {\cal A} (\omega) = \int d
\omega^\prime {\cal A}_\raw (\omega^\prime)$.  \\ (iii) \emph{Conservation of peak
  height:} for the choice $\gamma = \alpha/4$ (adopted henceforth) $L$ is
\emph{symmetric} under $\omega \leftrightarrow \omega^{\prime}$, so
that also $\int d \omega^\prime L(\omega, \omega^\prime) = 1$. This ensures that
the logarithmic broadening kernel maps a constant function onto
itself (if ${\cal A}_\raw (\omega^\prime) = {\cal A}_0$, then ${\cal
A}_\raw (\omega) = {\cal A}_0$), and thus does not change the height
of a peak whose width on a logarithmic scale is broader than
$\alpha$. (In this respect our $L$ differs from that of
\cite{SakaiShimizuKasuya89,Costi94,PetersPruschkeAnders06}.)

Since choice (iii) implies that our log-Gaussian Kernel, as function
of $\omega$, describes a peak asymmetric w.r.t. $\omega^\prime$ (shifted
by $\alpha/4$ on a log scale), on a linear $\omega$-scale the
broadened data is stretched relative to the raw data by factor
$e^{\alpha^2/4}$. This effect can be minimized by keeping $\alpha$
as small as possible. The smoothening of plain NRG data typically
requires $\alpha\sim 1/\sqrt{\Lambda}$ (e.g. $0.7$ for $\Lambda=2$).
However, smaller values (\eg\ $\alpha\leq0.3$ or even smaller) can
be achieved by using the ``$z$-trick''
\cite{YoshidaWhitakerOiveira90}: collect several (say $N_z$) sets of
discrete FDM-NRG data, each obtained from a different, slightly
shifted logarithmic grid $\{\Lambda^{-n -z} \}$ of discrete
frequencies, for $N_z$ different values of $z$ between $-0.5$ and
0.5, and average the results.  The hopping matrix elements along the
Wilson chain are recalculated for each $z$ by carefully
tridiagonalizing the underlying logarithmically discretized
Hamiltonian.

\emph{Self-energy representation:---} The accuracy of the results for
${\cal A}_\sigma (\omega)$ for the Anderson model can be improved by
expressing it in terms of the impurity self energy
\cite{BullaHewsonPruschke98}: first, note that ${\cal A}_\sigma
(\omega)= -\textrm{Im}[ {\cal G}^R_\sigma (\omega)]/\pi$, where ${\cal
  G}^R_\sigma (\omega)$ is the Fourier transform of ${\cal G}^R_\sigma
(t) = -i \theta (t) \langle \{c^{\vphantom{\dagger}}_{0\sigma} (t),
c_{0 \sigma}^{\dagger} \} \rangle _{T}$. An improved version for
${\cal G}^R_\sigma (\omega)$ can be obtained by expressing it as
\begin{eqnarray}
{\cal G}^\improved_\sigma (  \omega)
=\frac{1} {\omega-\Delta_\sigma (\omega)
-\Sigma_{\sigma}^U \left(  \omega\right)  } ,
\; \; \Sigma_{\sigma}^U \left(
\omega\right)  =U \frac{{\cal F}^R_\sigma \left(  \omega\right)  }
{{\cal G}^R_\sigma \left( \omega\right)}  . \nonumber
\end{eqnarray}
Here $\Delta_\sigma\left( \omega\right) $, the $U$-independent part of
the self energy which characterizes the level's broadening, can be
computed exactly, ${\cal G}^R_\sigma (\omega)$ is the standard
(``non-improved'') version of the correlator, and ${\cal F}^R_\sigma
\left( \omega\right)$ is the Fourier transform of $-i \theta (t)
\langle \{ [c^\dagger_{0 - \sigma} c^{\vphantom{\dagger}}_{0- \sigma}
c^{\vphantom{\dagger}}_{0\sigma}] (t), c_{0 \sigma}^{\dagger}\rangle
_{T}$. 

We calculate the imaginary parts of ${\cal G}^R_\sigma (\omega)$ and
${\cal F}^R_\sigma (\omega)$ using FDM-NRG from Lehmann
representations of the form (\ref{eq:correlation}), smoothen the
discrete data as described above, Kramers-Kronig transform the
smoothened results to obtain their real parts, and finally calculate
$\Sigma^U_{\sigma } (\omega)$.  Small wavy features in ${\cal
  G}^R_\sigma (\omega)$ and ${\cal F}^R_\sigma (\omega)$ that reflect
the logarithmic discretization grid largely cancel out in the ratio
$\Sigma^U_{\sigma } (\omega)$.  Thus, smooth results for ${\cal
  G}^\improved_\sigma ( \omega) $ and ${\cal A}^\improved_\sigma (
\omega) $ can be obtained using much less (or even no) $z$-trick
averaging, thus reducing the number of distinct FDM-NRG runs required
to get good results.  Moreover, since $\Sigma_{\sigma}^U (\omega\to
0)$ at $T=0$ is found to approach 0, the self-energy representation
also improves the accuracy with which the Friedel sum rule is
fulfilled.


\begin{thebibliography}{*}
\bibitem{app} The Appendix describes our smoothing procedure.
\end{thebibliography}
\end{document}